\documentstyle[11pt,newpasp,twoside,psfig]{article}
\begin{document}

\title{A First Close Look at the Balmer-edge Behavior of the Quasar Big
Blue Bump}
\author{Makoto Kishimoto}
\affil{Institute for Astronomy, University of Edinburgh,
Blackford Hill, Edinburgh EH9 3HJ, UK}

\author{Robert Antonucci and Omer Blaes}
\affil{Physics Department, University of California, Santa Barbara,
CA93106, USA}

\setcounter{page}{000}
\index{Kishimoto, M.}
\index{Antonucci, R.}
\index{Blaes, O.}

\begin{abstract}

We have found for the first time a Balmer edge feature in the Big Blue
Bump emission of a quasar. The feature is seen in the polarized flux
spectrum of the quasar, where all the emissions from outside the
nucleus are scraped off and removed. The existence of the Balmer-edge
absorption feature directly indicates that the Big Blue Bump is indeed
thermal and optically-thick.

\end{abstract}

\section{Introduction}

Among the various components of a quasar emission, the optical/UV
component, called Big Blue Bump (BBB), is energetically the most
dominant. The BBB is often assumed to be from optically-thick thermal
emission from an accretion flow, such as an accretion disk. However,
this fundamental issue (i.e. the emission mechanism of the BBB) is
actually very far from being solved, since observations are hardly
said to be well described by disk models.  Among several serious
problems are the continuum slope and apparent lack of continuum
edges. In order to explain the observed optical/UV slope
(e.g. $F_{\nu} \propto \nu^{-0.3}$, Francis et al. 1991), the disk has
to be rather cool.  However, naively, in a cool disk, its atmosphere
would show large continuum edges (but see below). Apparently, we do
not see such features.

\section{Balmer edge feature and our strategy}

In the sophisticated disk atmosphere models (e.g.  Hubeny et
al. 2000), when the spectrum is integrated over radii and smeared
through the relativistic Doppler shifts and gravitational redshifts,
the flux discontinuity at the Lyman limit is rather well
smeared. However, the {\it Balmer} edge, which arises further out in
radius where relativistic smearing is not too severe, {\it can still
be substantial}. The only problem is that it is very hard to be
observed due to the Balmer and FeII emission lines and nebular
continuum from the Broad Line Region (BLR) and outer regions (called
3000\AA\ bump).

However, we can remove all these unwanted emissions by taking a
polarized flux spectrum.  Many quasars are found to be polarized at $P
\sim 1-2$\%, and in many cases, emission lines (broad and narrow) are
unpolarized - the polarization is confined only in the continuum
(Antonucci 1988, Schmidt \& Smith 2000). This indicates that the
polarization mechanism resides interior to the BLR in these
cases. Then the polarized flux should show the intrinsic shape of the
BBB (except for a synchrotron possibility which is ruled out by the
result below at least for that object), revealing the underlying
behavior of the BBB.

\section{Results and Conclusions}

Figure~1 shows our recent Keck spectropolarimetry data of a
quasar. Thick line is the polarized flux spectrum, while the dotted
line is the total flux spectrum scaled to roughly match the polarized
flux at the red side. Emission lines are essentially all absent in the
polarized flux (see Fig.2 which shows that the polarization clearly
decreases at the line wavelengths).  The slope of the polarized flux
at the red side is roughly the same as that of the total
flux. However, we find that the slope changes at the bluer side of
$\sim$4000\AA\ in the rest frame, just where the 3000\AA\ bump starts
in the total flux (there might also be a possible up-turn at the bluer
side of $\sim$3600\AA).  This is an expected Balmer-edge absorption
feature, and directly indicates that the Big Blue Bump is indeed
thermal and optically thick, without involving any particular model of
the nucleus.

\bigskip

\unitlength 1cm

\begin{figure}
  \hbox to \textwidth{\hfil
    \begin{minipage}{7cm}
       \begin{picture}(5,5)
       \put(0,0){\psfig{figure=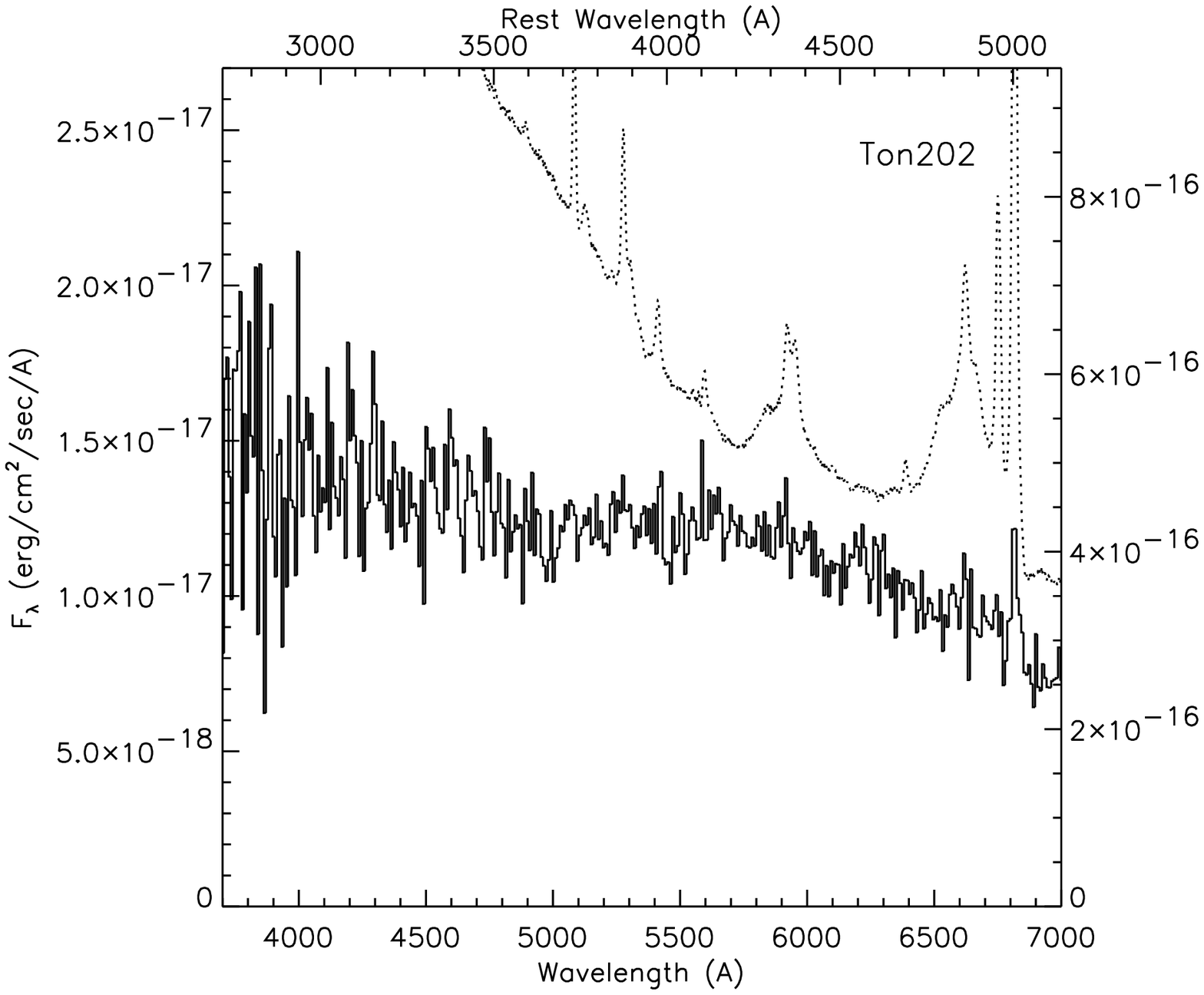,width=6cm}}
       \end{picture}
       \caption{{\footnotesize The polarized flux and total flux of
    Ton202, taken at Keck in May 2002.}}
    \end{minipage}
    \hfil
    \begin{minipage}{7cm}
       \begin{picture}(5,5)
       \put(0,0){\psfig{figure=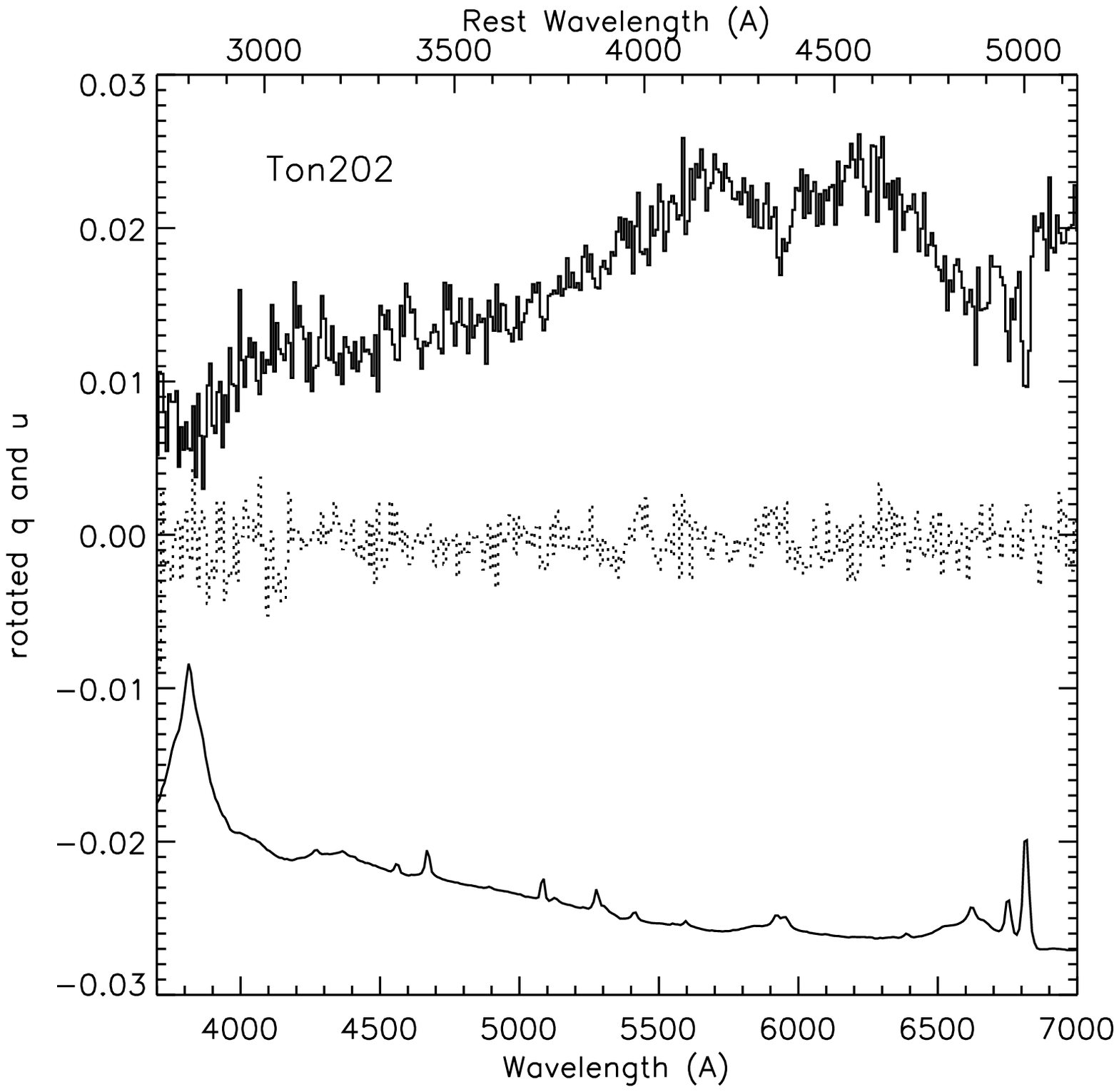,width=6cm}}
       \end{picture}
       \caption{{\footnotesize The normalized Stokes $q$
    and $u$, with the scaled total flux at the bottom for reference.}}
    \end{minipage}
    \hfil}
\end{figure}


\end{document}